\documentclass[conference,doublecolumn,letter]{IEEEtran}

\usepackage[english]{babel}
\usepackage[T1]{fontenc}
\usepackage[utf8]{inputenc}
\usepackage{subfig}
\usepackage{multirow}
\usepackage[font = scriptsize, labelfont = bf, labelsep = period, tableposition = top, singlelinecheck = false, justification = centering]{caption}
\usepackage{placeins,verbatim}
\usepackage{amssymb}
\usepackage{amsmath,eqnarray}
\usepackage{amsthm}
\usepackage{latexsym}
\usepackage{wrapfig}
\usepackage{float}
\usepackage{mathtools}
\usepackage{bm}
\usepackage{listings}
\usepackage{textcomp}
\usepackage{setspace}
\usepackage[dvipsnames]{xcolor}
\usepackage{etoolbox}
\usepackage{tikz}
\usetikzlibrary{shapes,snakes,calc}
\usepackage[innermargin=0.6in,outermargin=0.6in,top=0.5in,bottom=0.6in, textwidth=7.3in]{geometry}

\newtheorem{definition}{Definition}
\newtheorem{lemma}{Lemma}
\newtheorem{theorem}{Theorem}

\newtheorem{remark}{Remark}

\newif\ifcomment
\commenttrue
\commentfalse 

\newif\ifcommentLater
\commentLaterfalse

\newcommand{\ie}{\emph{i.e.}, }
\newcommand{\eg}{\emph{e.g.}, }

\definecolor{dartmouthgreen}{rgb}{0.05, 0.5, 0.06}

\makeatletter
\def\smalloverbrace#1{\mathop{\vbox{\m@th\ialign{##\crcr\noalign{\kern3\p@}%
  \tiny\downbracefill\crcr\noalign{\kern3\p@\nointerlineskip}%
  $\hfil\displaystyle{#1}\hfil$\crcr}}}\limits}
\makeatother

\makeatletter
\def\smallunderbrace#1{\mathop{\vtop{\m@th\ialign{##\crcr
   $\hfil\displaystyle{#1}\hfil$\crcr
   \noalign{\kern3\p@\nointerlineskip}%
   \tiny\upbracefill\crcr\noalign{\kern3\p@}}}}\limits}
\makeatother

\makeatletter
\renewcommand*\env@matrix[1][*\c@MaxMatrixCols c]{%
  \hskip -\arraycolsep
  \let\@ifnextchar\new@ifnextchar
  \array{#1}}
\makeatother

\AtBeginEnvironment{matrix}{\setlength{\arraycolsep}{5pt}}
\DeclareMathOperator{\lcm}{lcm}

\DeclareMathOperator{\RM}{RM}
\DeclareMathOperator{\RS}{RS}

\DeclareMathOperator{\evaluation}{eval}

\DeclareMathOperator{\dimension}{dim}
\DeclareMathOperator{\spacespan}{span}
\DeclareMathOperator{\support}{supp}

\def\={\coloneqq}							
\def\tpar#1{\left( #1 \right)}    			
\def\dim#1{\dimension \tpar{#1}}			
\def\eval#1{\evaluation \tpar{#1}}

\newcommand{\Co}{\mathcal{C}}
\newcommand{\D}{\mathcal{D}}
\newcommand{\CstarD}{\cC \star \cD}
\newcommand{\CstarE}{\cC \star E}
\newcommand{\CstarDperp}{\left(\CstarD \right)^\perp}
\newcommand{\CstarDstarE}{\cC^{\star D}_{\star E}}

\def\cC{\mathcal{C}}
\def\cD{\mathcal{D}}

\def\F{\mathbb{F}}

\newcommand{\ceil}[1]{\left\lceil #1 \right\rceil}

\newcounter{alp}
\newcounter{ara}
\newcounter{rom}

\title{Private Information Retrieval from Colluding and Byzantine Servers with Binary Reed--Muller Codes}

\author{
  \IEEEauthorblockN{Perttu Saarela, Matteo Allaix,~\IEEEmembership{Member,~IEEE}, Ragnar Freij-Hollanti, Camilla Hollanti,~\IEEEmembership{Member,~IEEE}}

	\IEEEauthorblockA{\small  Department of Mathematics and Systems Analysis\\ Aalto University, Finland\\ E-mails:
    \{perttu.e.saarela, matteo.allaix, ragnar.freij, camilla.hollanti\}@aalto.fi
	}
}
\IEEEoverridecommandlockouts

\begin{document}

\maketitle
\begin{abstract}
This paper is eligible for the Jack Keil Wolf ISIT Student Paper Award. 
In this work, a flexible and robust private information retrieval (PIR) scheme based on binary non-maximum distance separable (non-MDS) codes is considered. This combines previous works on PIR schemes based on transitive non-MDS codes on one hand, and PIR from MDS-coded Byzantine and non-responsive servers on the other hand. More specifically, a PIR scheme employing binary Reed--Muller (RM) codes tolerant to colluding, Byzantine, and non-responsive servers is constructed, and bounds for the achievable rates are derived under certain conditions. 
The construction of such schemes turns out to be much more involved than for MDS codes. Namely, the binary query vectors have to be selected with great care to hit the desired information sets, which is technically challenging as will be shown. 
\end{abstract}

\section{Introduction}

Private information retrieval (PIR) \cite{chor1995private,chor1998private} enables a user to download a data item from a database without revealing the identity of the retrieved item to the database owner (user privacy). If additionally the user is supposed to obtain no information about any file other than the requested file (server privacy), the problem is referred to as \emph{symmetric} PIR (SPIR) \cite{gertner1998spir}. In recent years, PIR has gained renewed interest in the setting of distributed storage systems (DSSs), where the servers are storing possibly large files and may \emph{collude}, \emph{i.e.}, exchange their obtained queries. To protect from data loss in the case of the failure of some number of servers, such systems commonly employ erasure-correcting codes, \emph{e.g.}, maximum distance separable (MDS) codes \cite{macwilliams1977theory}. Several constructions using different techniques have been proposed, \emph{e.g.}, \cite{tajeddine2018private,blackburn2020storage,zhou2020message,lavauzelle2019design}, among plenty of others. 

The capacity of (S)PIR has also been determined in a variety of settings, most often alongside with capacity-achieving constructions \cite{sun2017replicated,sun2018capacity,sun2017capacity,banawan2018capacity,wang2017linear,Banawan2019byzantine,Holzbaur2019ITW,wang2018epsilon,samy2021leaky}, but is still open in its full generality for coded and colluding servers \cite{freij2017private,Sun2018conjecture}. Significant progress toward the general coded colluded PIR capacity was recently made in \cite{holzbaur2022capacity}. 

One highly adaptable construction for PIR schemes is the \emph{star product} construction. It was first introduced in \cite{freij2017private} for coded storage and $t$-collusion and the best rate was shown to be achievable with generalized Reed--Solomon (GRS) codes.  It has been extended to, \eg binary and non-MDS codes \cite{freij2019transitivePIR}, regenerating codes \cite{lavauzelle2021MBR}, unresponsive and Byzantine servers \cite{tajeddine2019robustRS}, streaming  \cite{holzbaur2020streamingTIT}, random linear networks \cite{tajeddine2020networks}, and quantum PIR \cite{allaix2021capacity, song2020colluding}. Here, encouraged by the versatility of star product schemes and the computational efficiency of the binary field, we aim at constructing a robust binary scheme that is resistant toward colluding, non-responsive, and erroneous (Byzantine) servers. Working over the binary field is beneficial for real-life applications since it eases computation and thus some loss of rate might be acceptable.

Application of Reed--Muller (RM) codes in a PIR scheme is considered in \cite{freij2017RM} and the scheme was later improved in \cite{freij2019transitivePIR}. RM codes in PIR were also studied in \cite{kumar2017RM} where the model is based on the one in \cite{fazeli2015codes,yaakobi2015pir}. This model is very different from what we consider, aiming to minimize storage cost, while the PIR is done in a ``black box'' manner. Non-MDS codes are also studied in \cite{kumar2017private,Kumar2019nonMDS_TIT} where RM codes are also mentioned. Byzantine PIR schemes are considered in \cite{Banawan2019byzantine,tajeddine2019robustRS,wang2019byz,jia2020byz}. However, all of these schemes assume the usage of MDS codes. To the best of our knowledge, a Byzantine PIR scheme over the binary field will be presented for the first time in this paper.

In more detail, the contributions in this paper include:
\begin{itemize}
    \item Considering binary codes in a Byzantine setting for the first time.
    \item Combining the transitive capabilities of RM codes   \cite{freij2019transitivePIR} and the MDS coded Byzantine PIR scheme of \cite{tajeddine2019robustRS}. The combinations of the two schemes is surprisingly non-trivial due to the loss of the MDS property and restricting to a binary field. For non-MDS codes, we have to be more careful with the choice of servers from which data is downloaded in each iteration, since every combination of servers does not form an information set of the storage code. Moreover the use of binary polynomials limits our query construction since terms have to be reduced modulo the relation $z^\nu = z$ for $\nu \geq 1$, wherefore multiplication by a monomial is not a graded map. 
    \item Derivation of the best possible rate the given construction can achieve.
    \item Comparison of the new scheme to the Byzantine scheme with GRS codes.
    \item Discussion on the shortcomings of our scheme.
\end{itemize}
The main result of the paper is Theorem \ref{thm:robust} and its derivation can be found in Section \ref{sec:robust_RM_scheme}.

\section{Preliminaries} \label{sec:preliminaries}

We denote binary field by $\F_2$ and the set $\{1,2,\dots,n\}$ by $[n]$. The parameters of an $[n,k,d]$ linear code denote length, dimension, and minimum distance, respectively. A permutation of a codeword is the permutation of its components, \ie
for a permutation $\sigma$ in the symmetric group $\mathbb{S}_n$ and a codeword $c = \left[c_1,\dots,c_n \right]$ 
\begin{equation*}
    \sigma(c) = \left[c_{\sigma(1)},\dots,c_{\sigma(n)} \right].
\end{equation*} For a code $\Co$ and a permutation $\sigma \in \mathbb{S}_n$, if $\sigma(c) \in \Co$ for all $c \in \Co$, we say that $\sigma$ is an \emph{automorphism} of $\Co$. The automorphisms of $\Co$ form a group denoted by $\Gamma(\Co)$.
We say that a linear code is \emph{transitive} if the automorphism group of the code acts transitively on $[n]$, \ie for all $i,j \in [n]$ there exists $\sigma \in \Gamma(\Co)$ such that $\sigma(i) = j$. 

\begin{definition}
A PIR scheme is said to be $(t,a,b)$-\emph{robust} if it protects against $t$-collusion, $a$ unresponsive servers (erasures), and $b$ Byzantine servers (errors).
\end{definition} If $t, a$ and $b$ are clear from the context, the prefix $(t,a,b)$ will be dropped. The asymptotic capacity of $(t,a,b)$-robust PIR schemes was conjectured in \cite{tajeddine2019robustRS} to be \[ C = \frac{n-(k+t+a+2b-1)}{n} .\] The conjecture is shown to hold for strongly linear schemes  and for symmetric PIR (with ``matched'' randomness \cite{wang2017linear,wang2019mismatched}, to be precise) in \cite{Holzbaur2019ITW,holzbaur2022capacity}. We will use this conjecture as a reference point for our scheme.

\begin{definition}[Star Product]
Let $\Co$, $\D \subseteq \F_q^n$ be linear codes of length $n$ and let their codewords be of the form $\boldsymbol c = [c_1,\dots,c_n] \in \Co$ and $\boldsymbol d = [d_1, \dots, d_n] \in \D$. Then the star product of $\Co$ and $\D$ is
\begin{equation*}
    \Co \star \D = \spacespan\{ \boldsymbol c \star \boldsymbol d = [c_1d_1, \dots, c_nd_n] \,|\, \boldsymbol c \in \Co, \, \boldsymbol d \in \D\}.
\end{equation*}
\end{definition}

We will study DSSs encoded by Reed--Muller codes, which we will discuss next. For a more thorough exploration of Reed--Muller codes, we refer to \cite{RMtheory}.
\begin{definition}[Reed--Muller Code]
Let $r, m$ be positive integers such that $r \leq m$. Let $n = 2^m$ and $\{P_1, \dots, P_n\}$ be the set of all points in $\F_2^m$. Let $\F_2[\boldsymbol z ] = \F_2[z_1,\dots,z_m]$ be the polynomial ring of $m$ variables over $\F_2$. Then the \emph{$r$th order binary Reed-Muller code} is defined as the set
\begin{equation*}
    \RM(r,m) = \{ (f(P_1),\dots,f(P_n))\, |\, f\in \F_2[\boldsymbol z], \deg(f) \leq r \}.
\end{equation*}
\end{definition}
For a polynomial $f$, we denote its evaluation vector over all points $P_1,\dots,P_n$ by $\eval{f}=(f(P_1),\ldots,f(P_n))\in \F_2^{n}$.
The Reed--Muller code $\RM(r,m)$ has a generator matrix
\begin{equation}\label{eq:rm_generator}
    G_{\RM(r,m)} = \begin{bmatrix}
    \eval{1} \\
    \eval{z_1} \\
    \vdots\\
    \eval{z_m}\\
    \eval{z_1z_2}\\
    \vdots\\
    \eval{z_{m-r}z_{m-r+1}\dots z_m}
    \end{bmatrix}\,.
\end{equation}

Let us briefly summarize important features of Reed--Muller codes in the following two lemmas.

\begin{lemma} \label{rm_lemma}
Let $\Co = \RM(r,m)$ and $\D = \RM(r',m)$ such that $r, r' < m$ and $r + r' \leq m$. Then the following hold:
\begin{enumerate}
    \item $k = \dim{\Co} = \sum_{i = 0}^r \binom{m}{i}$
    \item Minimum distance $d = 2^{m-r}$
    \item $\Co^\perp = \RM(m-r-1, m)$
    \item $\Co \star \D = \RM(r+r', m$) 
\end{enumerate}
\end{lemma}

\begin{lemma}
Reed--Muller codes are transitive.
\end{lemma}
For the proofs of these lemmas see \cite{RMtheory}.

Reed--Muller codes exhibit a recursive structure:
\begin{equation*}
    \RM(r,m) = \begin{cases}
    \F_2^{2^m}, &\text{if}\; r=m\\
    (\boldsymbol u,\,\boldsymbol u+ \boldsymbol v),\quad &\text{if}\; r<m
    \end{cases},
\end{equation*} where $\boldsymbol u \in \RM(r,m-1)$ and $\boldsymbol v\in \RM(r-1,m-1)$. In particular, $\RM(r',m) \subseteq \RM(r,m)$ whenever $r' \leq r$. Furthermore, for any information set $\mathcal{I}$ of $\RM(r',m)$ there exists some information set of $\RM(r,m)$ containing $\mathcal{I}$.

\section{Robust scheme for Reed--Muller codes}\label{sec:robust_RM_scheme}\vspace{-2mm}
We begin by explicitly listing the requirements for a $(t,a,b)$-robust star product PIR scheme: \vspace{-1.5mm}
\begin{enumerate}
    \item A storage code $\Co$ and a retrieval code $\D$. 
    \item The star product $\Co \star \D$ and the dual of the star product $(\Co \star \D)^\perp$.
    \item To protect against $t$-collusion, we need $d_{\D^\perp} - 1 \geq t$.
    \item A symbol retriever vector $E$ such that $\Co^{\star \D}_{\star E} = \Co\star\D + \Co\star E$ is known.
    \item We need the minimum distance of $\Co^{\star \D}_{\star E}$ to be $\geq 2b + a + 1$.
    \item To be able to decode the response correctly, $\support(E)$ must contain an information set of $\Co$ and this information set must be contained in some information set of $\CstarDperp$. 
\end{enumerate}
Here the first three items were inferred from \cite{freij2017private}, the next two from \cite{tajeddine2019robustRS}, and the final item from \cite{freij2019transitivePIR}. Item 6 only becomes an issue when considering non-MDS codes. In \cite{freij2019transitivePIR} it is shown that transitivity of the codes $\Co$ and $\CstarD$ is sufficient for satisfying item 6.

To create a $(t,a,b)$-robust scheme with Reed--Muller codes, let us go through the previous list, step by step, and motivate the necessary conditions. Let us fix the parameters, $m,t,a$ and $b$. Our storage and retrieval codes will both be Reed--Muller codes, namely $\Co = \RM(r,m)$ and $\D = \RM(r',m)$, respectively. By Lemma \ref{rm_lemma}, it also follows that $\CstarD = \RM(r+r',m)$ and $\CstarDperp = \RM(m-r-r'-1,m)$. This checks off the first two items from the list. For item 3, we require that $2^{r'+1}-1 \geq t$, or equivalently $r' \geq \log(t+1) - 1$.

For items 4 and 5, we need to consider the choice of $E$. We need these criteria for the Byzantine and unresponsive servers. In order to correct the errors, we need to know which code we are working over. 
Much like in the scheme of \cite{tajeddine2019robustRS}, we will choose $E$ to be some polynomial evaluation. Furthermore, since Reed--Muller codes are algebraic-geometric codes, we choose the polynomials in such a way that $\CstarE$ is contained in some suitable RM code. 
More specifically, we choose $E$ such that $\CstarDstarE \subseteq \RM(r+r_e, m)$, where $r_e$ is given by item 5 on the list. The code $\RM(r+r_e, m)$ has minimum distance $d_{\Co^{\star D}_{\star E}} = 2^{m-r-r_e}$. Thus we have\vspace{-1mm}
\begin{equation}\label{eq:r_e}
    2^{m-r-r_e} \geq a+2b+1 \Rightarrow r_e \leq m - r - \log\left(a+2b+1\right).
\end{equation}
Additionally, to ensure that $\CstarE$ does not vanish when projected to $\CstarDperp$, we require that $r_e \geq r' + 1$. From these bounds, we get the feasibility criterion for our scheme. For the scheme to exists, we must have\vspace{-1mm} \begin{equation}\label{eq:m}
    m \geq r + \log\left((t+1)(a+2b+1)\right).
\end{equation} To use the least number of servers, we can give a strict equality for $m$ and $r_e$ by ceiling the logarithm, that is, by setting $m = r + \ceil{\log\left((t+1)(a+2b+1)\right)}$ and $r_e = m - r - \ceil{\log\left(a+2b+1\right)}$.

The terms of $\Co^{\star D}_{\star E}$ that survive the projection to $\CstarDperp$ are the terms whose degree are greater or equal to $r+r'+1$. Thus we have a range for the degrees of monomials whose coefficients we can expect to download, namely $[r+r'+1,r+r_e]$, and in much the same way, we have a range for the degrees of the polynomials $e(z)$, namely $[r'+1, r_e]$. Given that we can ``push'' information as coefficients of these monomials on each round, we can achieve the rate \vspace{-1mm}
\begin{equation*}
    \mathcal{R} = \frac{\sum_{i=r'+1}^{r_e}\binom{m}{r+i}}{2^m}
\end{equation*} whilst protecting against $\left(2^{r'+1}-1\right)$-collusion. The process of pushing data as coefficients of some monomials of certain degree
was done for the robust GRS scheme in \cite{tajeddine2019robustRS}. This time, however, working over a binary field and thus losing the powers of terms, we have to be more careful about how exactly this is done. 

Finally, for the last item we want to ensure that we can decode the response. This criterion is always met in the cases where $\CstarDstarE \subseteq \CstarDperp$. Due to the recursive definition, \ie the $(\boldsymbol u, \boldsymbol u + \boldsymbol v)$-construction, we know that any information set of $\CstarDstarE$, and therefore $\Co$, will be contained in some information set of $\CstarDperp$. This happens precisely when $r+r_e \leq m-r-r'-1$. This inequality can be further morphed into $\log\left(a+2b+1 \right) \geq r + \log\left(t+1\right)$. This tells us that the scheme works best when one wants to correct more errors than they want to protect against collusion, which may be a reasonable assumption in some applications. It was also shown \cite{freij2019transitivePIR} that even if more than $t$ servers collude, RM PIR schemes may not fail with relatively high probability. This is because every colluding set of servers might not include an information set of the retrieval code.  

\textbf{Encoding and Server setup: } Fix $r,a,b,t$ and choose $m,r',r_e$ as described above. Then we can download $\rho = \sum_{i=r'+1}^{r_e}\binom{m}{r+i}$ symbols per iteration. We choose the storage code to be an $[n,k,d]$ Reed--Muller code $\Co = \RM(r,m)$ with the parameters $[2^m, \sum_{i=0}^r\binom{m}{i}, 2^{m-r}]$. Choose $L$ and $S$ optimally by setting $L = \lcm(\rho,k)/k$ and $S = \lcm(\rho,k)/\rho$. $L$ and $S$ are free parameters that can be thought of as number of rows per file and number of iterations in the scheme, see \cite{freij2017private}. Let $X \in \F_2^{ML\times k}$ be our file system, where we have $M$ files with $L$ rows. The generator matrix $G_\Co$ will be of the same form as in Equation \eqref{eq:rm_generator}. For convenience we label the monomials, whose evaluations form the rows of $G_\Co$, as $f_\ell$ where $\ell \in [k]$. Hence the generator matrix will take the form
\begin{equation}
    G_{\Co} = 
    \begin{bmatrix}
    \eval{f_1} &
    \eval{f_2} &
    \cdots&
    \eval{f_k}
    \end{bmatrix}^T.
\end{equation} Let $Y = XG$ be our DSS and let $x^i_\ell = \begin{bmatrix} x^i_\ell(1) & \dots & x^i_\ell(k) \end{bmatrix}$ be a single row of a single file. Then  $x^i_\ell$ will be encoded as 
\begin{equation*}
    y^i = x^i_\ell \cdot G_\Co = \begin{bmatrix} x^i_\ell(1) & \dots & x^i_\ell(k) \end{bmatrix}\begin{bmatrix} f_1(P_1) & \dots & f_1(P_n)\\
         \vdots & \ddots & \vdots \\
         f_k(P_1) & \dots & f_k(P_n)\end{bmatrix}  
\end{equation*} and the information about $x^i_\ell$ stored on server $j$ in an encoded form is 
\begin{equation*}
    y^{i,\ell}_j =\sum_{\alpha = 1}^k x^i_\ell(\alpha)f_\alpha(P_j) \in \F_2,
\end{equation*} where $P_j \in \F_2^m$  is the $j$th evaluation point.
In other words, each row of a file defines a linear combination of the generator monomials which is then evaluated on the point $P_j$ and sent to the $j$th server. Let us define the storage polynomials $g^i_\ell$ as
\begin{equation*}
    g^i_\ell(\boldsymbol z) = \sum_{\alpha = 1}^k x^i_\ell(\alpha)f_\alpha(\boldsymbol z).
\end{equation*} 
Note that the degree of $g^i_\ell$ is at most $r$.
Our storage system then becomes
\begin{equation*}
    Y = \begin{bmatrix}
    g^1_1(P_1) &\hspace{-2mm} \dots & \hspace{-2mm} g^1_L(P_1) & \hspace{-2mm}\dots &\hspace{-2mm} g^M_1(P_1) &\hspace{-2mm} \dots &\hspace{-2mm} g^M_L(P_1)\\
    \vdots & \hspace{-2mm}\ddots & \hspace{-2mm}\vdots & \hspace{-2mm} &\hspace{-2mm} \vdots &\hspace{-2mm} \ddots & \hspace{-2mm}\vdots \\
    g^1_1(P_n) &\hspace{-2mm} \dots & \hspace{-2mm} g^1_L(P_n) &\hspace{-2mm} \dots &\hspace{-2mm} g^M_1(P_n) &\hspace{-2mm} \dots &\hspace{-2mm} g^M_L(P_n)\\
    \end{bmatrix}^T ,
\end{equation*} where the $j$th column corresponds to the data of the $j$th server. Retrieving the file $i$ is equivalent to finding the coefficients of the polynomials $g^i_1, \dots, g^i_L$.

\textbf{Query construction: } Let $\D = \RM(r', m)$ be the retrieval code. Then the star product code $\Co \star \D = \RM(r+r', m)$ is a Reed--Muller code with parameters $[2^m, \sum_{i=0}^{r+r'}\binom{m}{i}, 2^{m-r-r'}]$.

 Since Reed--Muller codes are polynomial codes, every codeword of $\D$ can be thought of as the evaluation of a polynomial $d^{\mu,(s)}_\ell(z)$ with degree $\deg\left(d^{\mu,(s)}_\ell(z) \right) \leq r'$, \ie $d^{\mu,(s)}_\ell = \eval{d^{\mu,(s)}_\ell(z)}$. We choose the matrix $E^{i,(s)} = \eval{e^{i,(s)}} \in \F_2^{L\times n}$ such that its rows are evaluations of polynomials $e_\ell^{i,(s)}(z)$ with degrees in the range $r'+1 \leq \deg\left(e_\ell^{i,(s)}(z) \right) \leq r_e$. See Remark \ref{remark:query_construction} for more details on choosing these polynomials. We define the query polynomials on iteration $s \in [S]$ for retrieving file $i$ as
\begin{equation*}
    q_\ell^{\mu,(s)}(z) = \begin{cases}
    d_\ell^{\mu,(s)}(z) + e_\ell^{i,(s)}(z)    & \text{if } \mu = i\\
    d_\ell^{\mu,(s)}(z)                                    & \text{if } \mu\neq i
    \end{cases}.
\end{equation*}
On round $s$ we send the query $q_j^{(s)}$ to server $j$, where 
\begin{align*}
    q_j^{(s)} = & \ \big[q_1^{1,(s)}(P_j), \dots, q_L^{1,(s)}(P_j), \dots,\\ 
    &\quad   q_1^{M,(s)}(P_j), \dots, q_L^{M,(s)}(P_j)\big]\,.
\end{align*}

\begin{remark}[Choosing the query polynomials]\label{remark:query_construction}
We will not give an explicit formula for the choice of the polynomials $e_\ell^{i,(s)}(z)$. 
However, we will give some general rules that should be taken into account when choosing them.

Firstly, it seems that ending up with linear combinations of desired symbols in the response is inevitable. This is not a problem if the queries are chosen such that these linear dependencies are solvable later on. Secondly, one should download the coefficients of higher degree terms first. Downloading lower degree terms will involve pushing higher order terms out of the range of error-correction. Thus, if one does not know the coefficients of the higher degree terms, they cannot be subtracted and thus the response cannot be error-corrected. For optimality, we suggest downloading higher degree terms from all stripes of the file before moving on to lower degrees.

Lastly, it is to be expected that one downloads symbols from only a portion of the stripes in a single round. In Example 1 we have $L=6$ stripes but sending the polynomials $e_\ell^{i,(s)}(z)$ to three stripes already uses up all of the polynomials whose coefficients will be downloaded. Thus sending polynomials to more stripes would diminish the number of explicitly downloaded symbols and increase the number of downloaded linear combinations. 
\end{remark}

\textbf{Responses: } After receiving the queries, the server responds with the inner product of its contents and the query, that is, $r_j^{i,(s)} = \left \langle y_j, q_j^{(s)} \right \rangle$. The total response polynomial without the inclusion of errors then becomes
\begin{equation*}
    r^{i,(s)} =  \sum_{\mu=1}^M\sum_{\ell = 1}^L d_\ell^{\mu,(s)}(z)g_\ell^\mu(z) + \sum_{\ell=1}^Le_\ell^{i,(s)}(z) g_\ell^i(z).
\end{equation*}
The term on the left has degree $\deg\left(d_\ell^{\mu,(s)}(z)g_\ell^\mu(z) \right) \leq r' + r$ and thus vanishes when projected to $\CstarDperp$. The term on the right has degree $ r+r'+1\leq\deg\left(e_\ell^{i,(s)}(z) g_\ell^i(z) \right) \leq r +r_e$ and thus that is also the degree of $r^{i,(s)}$. Hence, the response belongs to the code $\RM(r+r_e,m)$. By the choice of our fixed parameters,  we can error-correct up to $b$ errors and $a$ erasure symbols.

In full generality, the response polynomial can be written as 
\begin{equation*}
    r^{i,(s)} =  \delta^{i,(s)}(z) + h^{i,(s)}(z) + \gamma^{i,(s)}(z),
\end{equation*} where $$\delta^{i,(s)}(z) = \sum_{\mu=1}^M\sum_{\ell = 1}^L d_\ell^{\mu,(s)}(z)g_\ell^\mu(z)$$ contains the terms whose degree is less than $r+r'+1$ and can thus be considered as random interference. The term $\gamma^{i,(s)}(z)$ contains the terms whose coefficients are known from previous rounds. Note that $\gamma^{i,(s)}(z)$ might have degree larger than $r+r_e$ and must thus be subtracted before error-correction. The term $h^{i,(s)}(z)$ contains the new symbols we download on round $s$. Stated in a different way, $$h^{i,(s)}(z) + \gamma^{i,(s)}(z) = \sum_{\ell=1}^Le_\ell^{i,(s)}(z) g_\ell^i(z).$$

\textbf{Decoding: }  To decode the responses, the user first subtracts the known terms from the response, error-corrects it, and projects it to the code $(\CstarD)^\perp$ by left multiplying it with the generator matrix $H$ of $\CstarDperp$, 
\begin{align*}
    H r^{i,(s)} &= H\left(\eval{\delta^{(s)}(z)}+\eval{h^{i,(s)}(z)}\right)\\ &= H\eval{h^{i,(s)}(z)}.
\end{align*}

To recover the desired symbols, an information set of $\eval{h^{i,(s)}(z)}$ must be contained in an information set of $\CstarDperp$. This way, we can simply consider an invertible submatrix of $H$ given by the columns corresponding to the information set and recover the symbols by multiplying with the inverse. Given that $r+r_e \leq m-r-r'-1$ and the recursive nature of Reed--Muller codes, we can always find such an information set.

We have thus arrived at the following theorem.
\begin{theorem}\label{thm:robust} Let $\Co = \RM(r,m)$ be a storage code such that Equation \eqref{eq:m} holds and let $\D = \RM(r',m)$ be a retrieval code such that $r' \geq \log\left(t+1 \right) - 1$.  Given that $r_e$ is chosen as in Equation \eqref{eq:r_e} and $\log\left(a+2b+1 \right)-\log\left(t+1\right) \geq r$ holds, the above scheme  is $(t,a,b)$-robust. It can correct up to  $a$ erasures and $b$ errors as well as protect against $t$-collusion while achieving the rate
\begin{equation*}
    \mathcal{R} = \frac{\sum_{i=r'+1}^{r_e}\binom{m}{r+i}}{2^m}.
\end{equation*}
\end{theorem}

We conclude this section by giving a simple example using this scheme. However, working with Reed--Muller codes, even the simple examples grow large fairly quickly.

\vspace{1mm}
\hspace{-3.5mm}\textbf{Example 1.}  Consider the case where $m = 4, b=a=t=1$. Then our storage code is $\Co = \RM(1,4)$ and our retrieval code is $\D = \RM(0,4)$. We find that $r_e = m - \log(a+2b+1) - r = 1$ and hence \[\rho = \sum_{i=r'+1}^{r_e} \binom{m}{r+i} = \binom{4}{2} = 6.\] We set our free parameters as usual and get $L = 6$ and $S = 5$. The generator matrix for our storage code is 
\begin{equation*}
    G_\Co = \eval{\begin{bmatrix}
    1 & z_1 & z_2 & z_3 & z_4
    \end{bmatrix}^T}.
\end{equation*}
Hence each row of our file $x^i_\ell = (a^\ell_0, a^\ell_1, \dots, a^\ell_4)$ is encoded by the polynomial \[g^i_\ell(z) = a^\ell_0 + a^\ell_1z_1 + a^\ell_2z_2 + a^\ell_3z_3 + a^\ell_4z_4\] and the $j$th server has the evaluation $g^i_\ell(P_j)$.

\begin{figure*}[!ht]
\caption{On the left, PIR rate of robust RM scheme (solid) and robust RS scheme (dashed) in terms of length. Both schemes have a fixed code rate of $1/2$. Note that the rates match when there is no collusion and only one unresponsive server. In the middle, PIR rate in terms of storage code rate with fixed length $n=64$ and fixed parameters $t=a=1$. RM codes are represented by solid lines and RS codes by dashed lines. On the right, we fix $n=128$, $b=3$ and $a=1$.}
\label{fig:rate}
\centering
\begin{tabular}{c c c}
  \includegraphics[width=5.8cm]{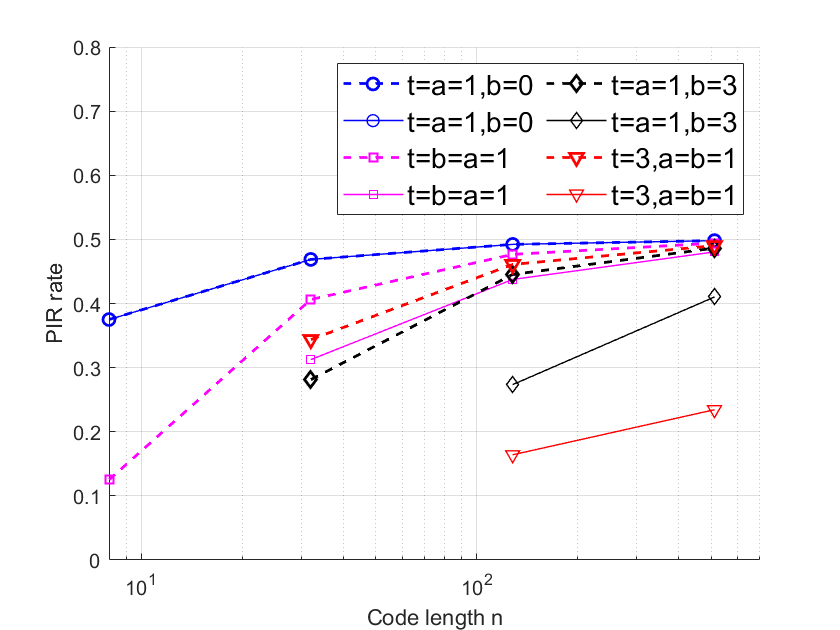} &
  \includegraphics[width=5.8cm]{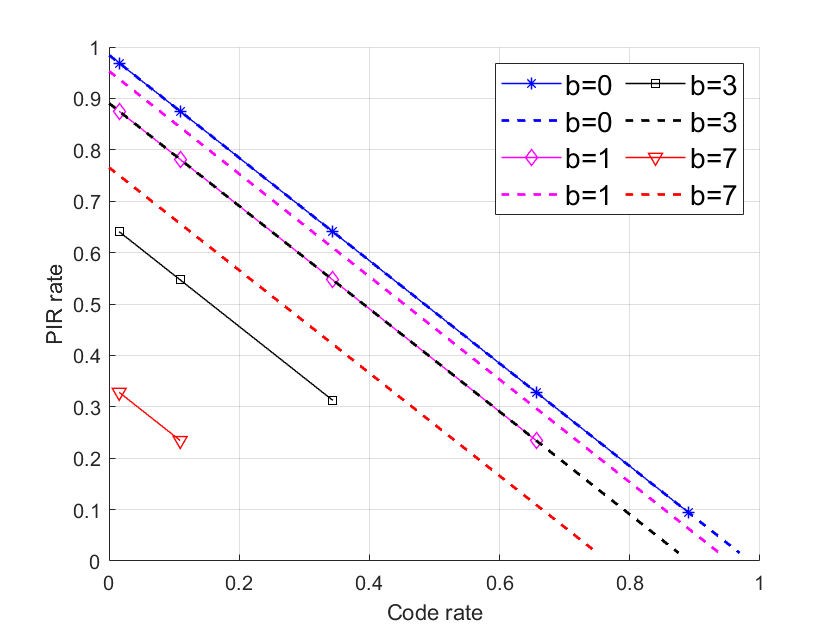} &
  \includegraphics[width=5.8cm]{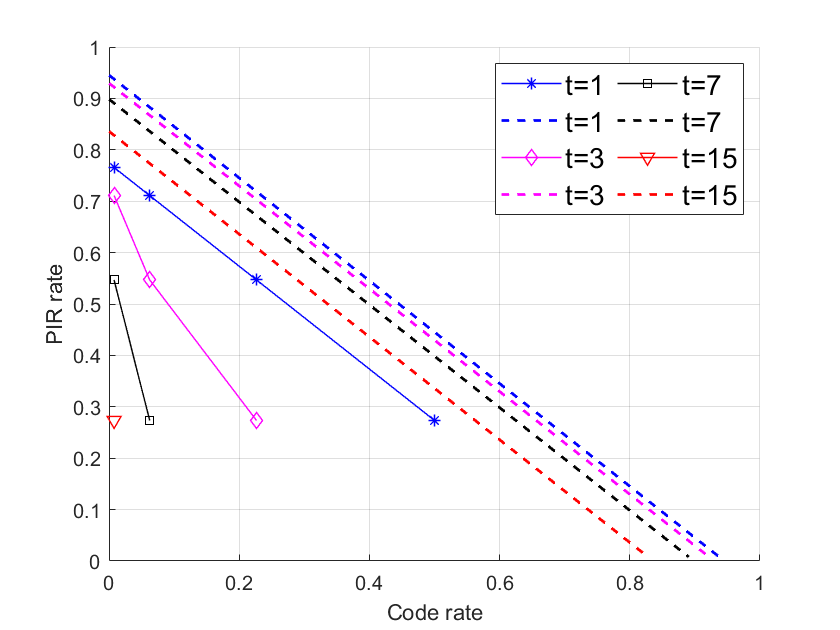} 
  \end{tabular}
\end{figure*}

We have $\rho = 6$ polynomials whose coefficients will be downloaded in each round. These polynomials are $A = \{z_1z_2 , z_1z_3 , z_1z_4 , z_2z_3 , z_2z_4 , z_3z_4 \}$. Considering that the polynomials $g^i_\ell$ are of degree one, we notice that we need three monomial terms in $e(z)$ so that $e(z)g^i_\ell$ contains polynomial terms in $A$. To avoid downloading the same symbol multiple times from a single row, we divide the monomial terms to three stripes. Thus on round $s=1$, we choose the polynomials
\begin{equation*}
    e^{i,(1)}(z) = \begin{bmatrix}
    z_1 & z_2 & z_3 & 0 & 0 & 0
    \end{bmatrix}^T.
\end{equation*}
Adding this to the random codewords of $\D$ and sending them to the servers, we get the response polynomial
\begin{align*}
    r^{i,(1)}(z) &= \delta^{(1)}(z) + a^1_4 z_1z_4 + a^2_4 z_2z_4 + a^3_4 z_3z_4+\\
    &+ (a^1_2+a^2_1) z_1z_2 + (a^1_3+a^3_1)z_1z_3 + (a^2_3+a^3_2)z_2z_3.
\end{align*}
After error-corrections and projection to $\CstarDperp$ we recover the symbols $a^1_4, a^2_4, a^3_4$ and three linear combinations, namely $(a^1_2+a^2_1), (a^1_3+a^3_1)$ and $(a^2_3+a^3_2)$.

In round $s=2$ we want to download some new symbols explicitly such that we end up recovering as many symbols from the linear combinations as possible. We choose the following polynomials
\begin{equation*}
    e^{i,(2)}(z) = \begin{bmatrix}
    z_2 & z_3 & z_4 & 0 & 0 & 0
    \end{bmatrix}^T
\end{equation*}
and get the corresponding response polynomial
\begin{align*}
    r^{i,(2)}(z) &= \delta^{(2)}(z) + a^1_1 z_1z_2 + a^2_1 z_1z_3 + a^3_1 z_1z_4 +\\
    &+ (a^1_3+a^2_2) z_2z_3 + (a^1_4+a^3_2)z_2z_4 + (a^2_4+a^3_3)z_3z_4.
\end{align*}
After error-corrections and projections we again recover three symbols explicitly, namely $a^1_1, a^2_1, a^3_1$, and three linear combinations of symbols, which are $(a^1_3+a^2_2), (a^1_4+a^3_2)$ and $(a^2_4+a^3_3)$. This time, however, we already know some symbols and can use the known symbols and linear combinations to recover new symbols. Working our way through the linear combinations, we recover all coefficients of the degree $1$ terms of the first three rows bit by bit.

For rounds $s=3,4$, we repeat the same process as in rounds $s=1,2$, but this time sending the polynomial evaluations to the rows $\ell = 4,5,6$. That is, we have the queries \begin{align*}
    e^{i,(3)}(z) &= \begin{bmatrix}
    0&0&0& z_1 & z_2 & z_3 
    \end{bmatrix}^T\\
    e^{i,(4)}(z) &= \begin{bmatrix}
    0&0&0& z_2 & z_3 & z_4
    \end{bmatrix}^T
\end{align*} and the corresponding responses
\begin{align*}
    r^{i,(3)}(z) &= \delta^{(3)}(z) + a^4_4 z_1z_4 + a^5_4 z_2z_4 + a^6_4 z_3z_4 +\\
    &+ (a^4_2+a^5_1) z_1z_2 + (a^4_3+a^6_1)z_1z_3 + (a^5_3+a^6_2)z_2z_3\\
    r^{i,(4)}(z) &= \delta^{(4)}(z) + a^4_1 z_1z_2 + a^5_1 z_1z_3 + a^6_1 z_1z_4 +\\
    &+ (a^4_3+a^5_2) z_2z_3 + (a^4_4+a^6_2)z_2z_4 + (a^5_4+a^6_3)z_3z_4.
\end{align*}
In doing so, we again recover the coefficients of all degree $1$ terms of the last three rows.

Finally, on round $s=5$ we can download the coefficients of the degree $0$ terms by sending each of our degree $2$ query polynomials to the different rows. This entails that our response will have degree $3$ terms which we cannot error-correct. However, we know what those terms are and the corresponding coefficients from previous rounds. Thus we can subtract the higher degree terms before error-correction.

We have the query polynomial
\begin{equation*}
    e^{i,(5)}(z) = \begin{bmatrix}
    z_1z_2 & z_1z_3 & z_1z_4 & z_2z_3 & z_2z_4 & z_3z_4
    \end{bmatrix}^T
\end{equation*}
and the corresponding response polynomial
\begin{align*}
    r^{i,(5)}(z) = \delta^{(5)}(z) &+ a^1_0 z_1z_2 + a^2_0 z_1z_3 + a^3_0 z_1z_4 +\\
    &+ a^4_0 z_2z_3 + a^5_0 z_2z_4 + a^6_0 z_3z_4 + \gamma(z),
\end{align*} where $\gamma(z)$ is the polynomial containing all of the degree $3$ terms and the degree $2$ terms with known coefficients. Note that the choice of which degree two monomial is sent to which stripe is irrelevant.

We conclude the example by tabulating which coefficients were recovered on which round in Table \ref{tab:byz_rm_ex}. Some coefficients are not downloaded explicitly but are downloaded in a linear combination. In that case we will mark down the iteration when the coefficient is recovered, \ie can be solved for, not when it is downloaded.
\begin{table}[b]
    \centering
    \begin{tabular}{l||c|c|c|c|c}
        $_\ell$ $^{\kappa}$ & 0 & 1 & 2 & 3 & 4  \\
         \hline
        1 & \textcolor{red}{5} & \textcolor{violet}{2} & \textcolor{violet}{2} & \textcolor{violet}{2} & \textcolor{blue}{1} \\
        2 & \textcolor{red}{5} &\textcolor{violet}{2} &\textcolor{violet}{2} &\textcolor{violet}{2} &\textcolor{blue}{1} \\
        3 & \textcolor{red}{5} &\textcolor{violet}{2} &\textcolor{violet}{2} &\textcolor{violet}{2} &\textcolor{blue}{1}\\
        4 & \textcolor{red}{5} & \textcolor{olive}{4} & \textcolor{olive}{4} & \textcolor{olive}{4} & \textcolor{orange}{3} \\
        5 & \textcolor{red}{5} &\textcolor{olive}{4} &\textcolor{olive}{4} &\textcolor{olive}{4} &\textcolor{orange}{3} \\
        6 & \textcolor{red}{5} &\textcolor{olive}{4} &\textcolor{olive}{4} &\textcolor{olive}{4} &\textcolor{orange}{3}\\
    \end{tabular}
    \caption{Table of downloaded coefficients $a_\ell^\kappa$ and in which iteration they are recovered. Rows correspond to stripes of a file, columns to terms.}
    \label{tab:byz_rm_ex}
\end{table}
With this scheme, we achieve the rate
\begin{equation*}
    \mathcal{R}_{\RM} = \frac{Lk}{Sn} = \frac{\rho}{n} = \frac{6}{2^4} = \frac{3}{8}.
\end{equation*} 
The robust scheme with Reed--Solomon codes would give the rate 
\begin{align*}
    \mathcal{R}_{\RS} &= \frac{n-k-t-2b-a+1}{n}\\  &= \frac{16-5-1-2-1+1}{16} = \frac{1}{2}=\frac{4}{8}, 
\end{align*} which is only slightly better than with our binary scheme. However, the RS scheme would require a field size of at least $q \geq 16$ whereas the RM scheme works over $\F_2$. 

Further comparison between the robust RS scheme and the robust RM scheme can be seen in Figure \ref{fig:rate}. The plots demonstrate that the RM scheme works better with more errors and less collusion. However, unlike MDS-codes, RM codes can still protect against more than $t$ collusions with positive probability, which decreases as the number of collusions beyond $t$ grows, see \cite{freij2017RM} for more details.  Moreover, we see that rate is traded off for binary computation as was expected.  
For the left and the middle figures, in the plots for parameters $t=a=1, b=0$ the two schemes have equivalent rate. That is due the the schemes reducing to the same replication scheme. Note that the parameters $t,a,b$ are chosen conveniently such that the logarithm of Equation \eqref{eq:m} is an integer. One would have to ceil the logarithm for non-integer values. Thus choosing $a+2b+1$ and $t+1$ to be powers of two is optimal for the RM scheme. 

\section{Conclusions and future work} 

Future work includes making the query construction of the given new scheme explicit for any given parameters. This might not be entirely trivial but Remark \ref{remark:query_construction} and the above example give some insights on how to do this. Especially for the case where $r_e = r' + 1$, \ie there is only a single polynomial degree for which coefficients can be downloaded, the generalization should be straightforward.

An issue with Reed--Muller codes is their exponential length in terms of the parameter $m$. Hence, other shorter transitive or algebraic-geometric codes could be applied instead in a similar fashion. A promising family of transitive codes are cyclic codes which are closed under taking dual and star products. A challenge with cyclic codes, however, is that there is no closed formula for their minimum distance. PIR with cyclic codes is mentioned in \cite{freij2019transitivePIR,kumar2018nonMDS_ITW,Kumar2019nonMDS_TIT} but no explicit robust scheme has been given for them as of now. 

Another potentially fruitful path for robust PIR schemes might be via the usage of Goppa codes which are a family of algebraic-geometric codes. They have great error-correction capabilities, also work in binary, and are closed under star product for certain parameters. A challenge with Goppa codes is that there is no known explicit formula for the parameters of their dual codes so rates can only be bounded.

Reed--Muller codes have the interesting property of being weakly self-dual when the dimension is greater or equal than half the length, \ie their dual codes are subsets of the respective original codes. By employing this property and quantum stabilizer formalism \cite{Gottesman97}, one can setup a quantum PIR protocol with classical storage, classical queries, and quantum responses in order to double the rate of the classical counterpart in a similar way to the one described in \cite{allaix2021capacity}.

\section*{Acknowledgments}
This work was supported by the Academy of Finland, under Grants No. 318937 and 336005.


\newpage

\bibliographystyle{IEEEtran}
\bibliography{main}

\end{document}